\documentclass[aps,prl,twocolumn,showpacs,superscriptaddress,groupedaddress]{revtex4}
\usepackage{graphicx}  
\usepackage{bm}        
\usepackage{amssymb}   
\usepackage{amsmath}
\usepackage{color}

\begin{document}

\title{Entropy production and  fluctuation theorems for Langevin processes under continuous non-Markovian feedback control}
\author{T. Munakata}
\affiliation{Department of Applied Mathematics and Physics, 
Graduate School of Informatics, Kyoto University, Kyoto 606-8501, Japan} 
\email{tmmm3rtk@hb.tp1.jp}
\author{M.L. Rosinberg}
\affiliation{Laboratoire de Physique Th\'eorique de la Mati\`ere Condens\'ee, Universit\'e Pierre et Marie Curie, CNRS UMR 7600,\\ 4 place Jussieu, 75252 Paris Cedex 05, France}
\email{mlr@lptmc.jussieu.fr}

\begin{abstract}
Continuous feedback control of Langevin processes may be non-Markovian due to a time lag between the measurement and the control action. We show that this requires to modify the basic relation between dissipation and time-reversal and to include a contribution arising from the non-causal character of the reverse process. We then propose a new definition of the quantity measuring the irreversibility of a path in a nonequilibrium stationary state, which can be also regarded as the trajectory-dependent total entropy production. This leads to an extension of the second law which takes a simple form in the long-time limit. As an illustration, we apply the general approach to linear systems which are both analytically tractable and experimentally relevant.
\end{abstract}

\maketitle

{\it Introduction.--} As famously illustrated by Maxwell's demon thought experiment\cite{MNV2009}, entropy production (EP) in small thermodynamic systems can be reduced by the intervention of an external agent who possesses some information about the microstates. Recent years have seen a renewed interest for this idea due to the advances in the manipulation of mesoscopic objects and to a better understanding of  the intimate relationship between EP and time asymmetry at the microscopic level\cite{C1999}. The ultimate goal of these investigations is to develop  a `thermodynamics of feedback', relating information and dissipation\cite{CF2009,SU2012}.
 
With this goal in mind, we focus in this Letter on classical stochastic systems described by a Langevin dynamics and submitted to a continuous, non-Markovian feedback control.  The non-Markovian character results from a time lag between the signal detection and the control action, which is an ubiquitous feature in biological systems\cite{BR2000} and  also plays an important role in many experimental setups  ({\it e.g.} laser networks\cite{SGMF2013}). Because of memory effects, the conventional approach of stochastic thermodynamics\cite{J2011} is not applicable to such systems, and even the basic identity (the so-called local detailed balance condition) which is at the heart of fluctuation relations\cite{C1999} needs to be modified. Indeed, in order to relate the heat dissipated along an individual  trajectory to the statistical weights of the trajectory and its time-reversal, causality must be artificially broken  in the backward process, giving rise to a specific ``Jacobian" contribution.   
Such effect went unnoticed in previous theoretical studies which mainly focused on discrete feedback protocols in which the controller acts at predetermined times.  In this case, the reverse process is physically realizable\cite{TSUMS2010}, which is not possible when the feedback is applied continuously. This prompts us to propose  a new  definition of the fluctuating entropy production in a nonequilibrium stationary state (NESS), which in turn leads to a generalization of the second law. We  illustrate this  general approach by a detailed analytical and numerical study of linear systems. 
Note that the present study is restricted to the case of a deterministic  ({\it i.e.} error-free) feedback control. Noise and measurement errors are known to reduce the achievable entropy reduction\cite{CF2009}.

{\it Dissipation and time-reversal.--} Without loss of generality, we  consider the one-dimensional motion of a  Brownian particle (or ``system'') in contact with a heat bath in equilibrium at inverse temperature $\beta$ (Boltzmann constant is set to $1$ hereafter). The dynamics is  described by a second-order Langevin equation with additive noise
\begin{align}
\label{EqL1}
m\ddot x+\gamma \dot x-F(x)-F_{fb}(t)=\xi(t)
\end{align}
where $m$ is a mass, $\gamma$ is a friction coefficient, $F(x)=-dU(x)/dx$ is  a conservative force, and  $\xi(t)$ is a delta-correlated white noise with variance $2\beta^{-1}\gamma$ (for simplicity, a memory-less friction is assumed, but the formalism can be generalized to a non-Markovian bath, as considered in previous studies\cite{ZBCK2005,OO2007,ABC2010}). $F_{fb}(t)$ is the feedback control force determined by the measurement outcomes and which generally depends on the microscopic trajectory of the system in phase space up to time $t$. It may be for instance proportional to the position $x$ at time $t-\tau$, where $\tau$ is the delay (see Eq. (\ref{EqDL1}) below), or to the velocity $\dot x$, as illustrated  by Eq. (\ref{Eqvelo}) where $\tau$ is  the relaxation time of the feedback mechanism\cite{MR2013}. This latter case  is a non-Markovian generalization of the  model studied in \cite{KQ2004,MR2012}  which describes feedback cooling (or cold damping) experimental setups\cite{PZ2012}.

In the normal operating regime of a continuous feedback control, the system settles into a NESS in which heat is permanently exchanged with the thermal environment (the stability of the NESS depends on the various parameters that specify  the dynamics, {\it e.g.} the  delay $\tau$). Within the framework of stochastic energetics\cite{Seki1997}, the heat dissipated along an individual path ${\bf {\bf X}}\equiv \{x_t,\dot x_t\}$ during the time interval $[-T,T]$ is then defined as
\begin{align}
\label{EqDeltasm}
&q[{\bf X},{\bf X}_-]\equiv \int_{-T}^{T} dt\:[\gamma \dot x_t-\sqrt{2\gamma \beta ^{-1}}\xi_t]\:  \dot x_t \nonumber\\
&=-\int_{-T}^{T} dt\: \Big[m \ddot x_t -F(x_t)-F_{fb}[{\bf X},{\bf X_-}]\Big] \: \dot x_t  \ ,
\end{align}
where ${\bf X_-}$ denotes the path for $t\le -T$ (we now make explicit the fact that $F_{fb}(t)$ depends on both  ${\bf X}$ and ${\bf X_-}$). 

As in the case of  Markov processes, we seek to relate $q[{\bf X},{\bf X}_-]$ to the time reversibility of the trajectories, so we  consider the probability of observing ${\bf X}$ for a given initial state ${\bf x}_i\equiv (x_{-T},\dot x_{-T})$ and  a given past trajectory ${\bf X_-}$\cite{note1,MRT2014}. This  probability is  determined by the noise history in the time interval $[-T,T]$ and  given by
\begin{align}
\label{Eqpath0}
&{\cal P}[{\bf X}\vert {\bf x}_i,{\bf X_-}]\propto \big \vert {\cal J} \big \vert \: e^{-\beta \int_{-T}^{T}dt \:  {\cal S}[{\bf X},{\bf X_-}]} \ ,
\end{align}
where ${\cal S}[{\bf X},{\bf X_-}]$ is a generalized  Onsager-Machlup action functional\cite{OM1953}, 
\begin{align}
\label{Eqaction0}
 {\cal S}[{\bf X},{\bf X_-}]&=\frac{1}{4\gamma}\Big[m\ddot x_t+\gamma \dot x_t-F(x_t)-F_{fb}[{\bf X},{\bf X_-}]\Big]^2\ ,
\end{align}
and  ${\cal J}$ is the Jacobian of the transformation $\xi(t)\rightarrow x(t)$ for $t\in [-T,T]$. Eq. (\ref{Eqpath0}) can be made rigorous by discretizing the Langevin dynamics, as done for instance in \cite{ABC2010} (in particular,  there is no need  to specify the interpretation of the stochastic calculus as long as $m\ne 0$).  Due to causality, the  Jacobian matrix is lower triangular, so that ${\cal J}$ is  a path-independent positive quantity that can be included in the prefactor\cite{note2}.

We now replace the whole trajectory, including ${\bf X_-}$, by its time-reversed image $\{ x^\dag(t), \dot x^\dag (t)\}=\{x(-t),-\dot x(-t)\}$ and consider the probability ${\cal P}[{\bf  X^\dag}\vert {\bf x}_i^\dag, {\bf  X_-^\dag}]$ of observing the reversed path $ {\bf  X^\dag}$, given the path $ {\bf  X_-^\dag}$ for $t\ge T$  and the initial state $ {\bf x}_i^\dag=(x_T,-\dot x_T)$. It is readily seen that in order to relate $q[{\bf X},{\bf X}_-]$ to the probabilities of ${\bf X}$ and ${\bf X}^\dag$, one must define a new feedback force $\tilde F_{fb}$ such that $\tilde F_{fb}[{\bf  X^\dag},{\bf  X_-^\dag}]_{t\rightarrow -t}=F_{fb}[{\bf X},{\bf X_-}]$. (In the same vein,  the driving protocol must be reversed in the case of a discrete feedback.)
Consider for instance a time-delayed feedback $F_{fb}\propto x(t-\tau)$. Then $\tau$ must be changed into $-\tau$ in order to recover the original force. Similarly, in the case of an exponential memory kernel,   $F_{fb}\propto \frac{1}{\tau}\int_{t_0}^{t}ds \: e^{-\frac{t-s}{\tau}}x(s)$, one must change $\tau$ into $-\tau$ and $t_0$ into $-t_0$.  For a velocity-dependent feedback like in Eq. (\ref{Eqvelo}), one must also change $\gamma'$ into $-\gamma'$. More generally, such changes define a ``conjugate" dynamics, hereafter denoted by the tilde symbol $(\sim)$. This dynamics is non-causal and  does not correspond to any physical  process, but the conditional probability
\begin{align}
\label{Eqpath1}
\tilde{\cal P}[{\bf  X^\dag}\vert {\bf x}_i^\dag, {\bf  X_-^\dag}]\propto \big \vert \tilde {\cal J}[{\bf X}]\big\vert \: e^{-\beta \int_{-T}^{T}dt \: \tilde {\cal S}[{\bf  X^\dag},{\bf  X_-^\dag}]}
\end{align}
with 
\begin{align}
\label{Eqaction1}
\tilde  {\cal S}[{\bf  X^\dag},{\bf  X_-^\dag}]=\frac{1}{4\gamma}\Big[m\ddot x_t-\gamma \dot x_t-F(x_t)-\tilde F_{fb}[{\bf  X^\dag},{\bf  X_-^\dag}]_{t\rightarrow -t}\Big]^2 
\end{align}
 is  a well-defined mathematical object. On the other hand, non-causality makes the Jacobian matrix no longer lower triangular and $\tilde{\cal J}[{\bf X}]$ is in general a nontrivial (positive) functional of the path (see Eq. (\ref{Eqexp}) below). Taking the ratio of ${\cal P}[{\bf X}\vert {\bf x}_i,{\bf X_-}]$ and $\tilde{\cal P}[{\bf  X^\dag}\vert {\bf x}_i^\dag, {\bf  X_-^\dag}]$  then  leads to our first main result 
\begin{align}
\label{Eqratio}
\frac{{\cal P}[{\bf X}\vert {\bf x}_i, {\bf X_-}]}{\tilde {\cal P}[ {\bf X}^\dag\vert {\bf x}_i^\dag, {\bf X}_-^\dag]}=\frac{{\cal J}}{\tilde {\cal J}[{\bf X}]} \exp \big\{\beta \: q[{\bf X},{\bf X_-}] \big\} \ ,
\end{align}
which generalizes the familiar identity relating dissipation to time reversal\cite{C1999}.  The two signatures of non-Markovianity  are (i) the functional dependence on the past trajectory, and (ii) the presence of the ratio ${\cal J}/\tilde {\cal J}[{\bf X}]$ due to the non-causal character of the  dynamics $\sim$. 

{\it Entropy production (EP).--} As in the Markovian case,  the left-hand side of Eq. (\ref{Eqratio}) may be combined with normalized distributions  ${\cal P}_{st}[ {\bf x}_i, {\bf X_-}]$ and ${\cal P}_{st}[{\bf x}_i^\dag, {\bf X}_-^\dag]$ in order  to define unconditional path weights. We thus introduce the quantity $R[{\bf X},{\bf X_-}]\equiv \Delta s_{m}[{\bf X},{\bf X_-}]-\ln \tilde {\cal J}[{\bf X}]/{\cal J}+\ln {\cal P}_{st}[ {\bf x}_i, {\bf X_-}]/{\cal P}_{st}[{\bf x}_i^\dag, {\bf X}_-^\dag] $, where $\Delta s_{m}[{\bf X},{\bf X_-}]\equiv \beta q[{\bf X},{\bf X_-}]$ is  the change in the entropy of the medium. By construction, $R[{\bf X},{\bf X_-}]$ satisfies the integral fluctuation theorem (IFT) 
\begin{align}
\label{EqIFT0}
\langle e^{-R[{\bf X},{\bf X_-}]}\rangle_{st}=1 
\end{align}
where  $\langle...\rangle_{st}$ denotes an average over all paths ${\bf X}$ and ${\bf X_-}$ weighted by the stationary probability ${\cal P}_{st}[{\bf X},{\bf X_-}]$. It is  worth noting that $R[{\bf X},{\bf X_-}]$ can also be expressed  as 
\begin{align}
\label{EqR}
R[{\bf X},{\bf X_-}]&=\Delta s_{tot}[{\bf X},{\bf X_-}]-\ln \tilde {\cal J}[{\bf X}]/{\cal J} \nonumber\\
& -\Delta {\cal I}[{\bf X}_-,{\bf x}_i,{\bf x}_i^\dag]+\ln {\cal P}_{st}[{\bf X_-}]/{\cal P}_{st}[ {\bf X}_-^\dag]
\end{align}
where  $\Delta s_{tot}[{\bf X},{\bf X_-}]\equiv \Delta s_m[{\bf X},{\bf X_-}]+\ln p_{st}({\bf x}_i)/p_{st}({\bf x}_i^\dag)$ is a  ``Markovian-like''  contribution\cite{J2011} and $\Delta {\cal I}={\cal I}[{\bf x}_i^\dag: {\bf X}_-^\dag]-{\cal I}[{\bf x}_i:{\bf X}_-]=\ln {\cal P}_{st}[{\bf x}_i^\dag\vert {\bf X_-}^\dag]/p_{st}({\bf x}_i^\dag)-\ln {\cal P}_{st}[ {\bf x}_i\vert {\bf X_-}]/p_{st}({\bf x}_i)$ describes memory effects not contained in $\Delta s_{tot}[{\bf X},{\bf X_-}]$ (here, ${\cal I}$ is a fluctuating mutual information).
 A drawback, however,  is that  $R[{\bf X},{\bf X_-}]$  do not vanish when the feedback control is switched off and the system goes back to equilibrium  (whereas $\Delta s_{tot}=0$). This problem is  cured by considering the coarse-grained  functional $R_{cg}[{\bf X}]=-\ln \int {\cal D} {\bf X}_- {\cal P}[{\bf X}_-\vert {\bf X}] \: e^{-R[{\bf X},{\bf X_-}]}$  which, from the definition of $R[{\bf X},{\bf X_-}]$, simply reads 
\begin{align}
R_{cg}[{\bf X}]\equiv \ln \frac{{\cal P}_{st}[{\bf X}]}{\tilde {\cal P}_{st}[ {\bf X}^\dag]}\ ,
\end{align}
where $\tilde {\cal P}_{st}[{\bf X}^\dag]\equiv \int {\cal D}{\bf X}_- \tilde{\cal P}[ {\bf X}^\dag\vert {\bf x}_i^\dag, { {\bf X}}_-^\dag]{\cal P}_{st}[{\bf x}_i^\dag, { {\bf X}}_-^\dag]$\cite{note3}. By construction $R_{cg}[{\bf X}]$ obeys the IFT, and its average
\begin{align}
\label{EqRcgav}
\langle R_{cg}[{\bf X}]\rangle_{st}=\int {\cal D}{\bf X}\: {\cal P}_{st}[{\bf X}] \ln \frac{{\cal P}_{st}[{\bf X}]}{\tilde {\cal P}_{st}[ {\bf X}^\dag]}
\end{align}
is the Kullback-Leibler divergence $D({\cal P}_{st}\vert\vert \tilde {\cal P}_{st})$ between the distributions ${\cal P}_{st}$ and $\tilde {\cal P}_{st}$. This quantity is always non-negative, which suggests that $R_{cg}[{\bf X}]$ properly describes the overall EP along the trajectory ${\bf X}$ as a measure of the  irreversibility of the non-Markovian  stationary process. In particular, $R_{cg}[{\bf X}]$  does not vanish when ${\cal P}_{st}[{\bf X}]={\cal P}_{st}[{\bf X}^\dag]$, which occurs when  all forces are linear (see below). 

{\it Asymptotic relations.--} $R_{cg}[{\bf X}]$, however, is a complicated functional of the path (see  \cite{MRT2014} for explicit calculations). On the other hand, its average has a simple expression when the observation time becomes much larger than the time constant characterizing the non-Markovian feedback (we here assume that the correlation to the past is finite or decreases rapidly with time, {\it e.g.} exponentially). The dependence on the past trajectory can then be neglected, as well as  the ``border" terms which are non extensive in time. This leads to the asymptotic equality
\begin{align}
\label{Eqasym}
\langle R_{cg}[{\bf X}]\rangle_{st}\sim \langle\Delta s_m[{\bf X}]\rangle_{st}-\langle\ln  \frac{\tilde {\cal J}[{\bf X}]}{{\cal J}} \rangle_{st}\ ,
\end{align}
which can be rewritten as $\dot R_{cg}=\dot S_m-\dot S_{{\cal J}}$ by defining the rates $\dot R_{cg}=\lim_{T\rightarrow \infty}\frac{1}{2T} \langle R_{cg}[{\bf X}]\rangle_{st}$, $\dot S_m=\frac{1}{2T} \langle s_m[{\bf X}]\rangle_{st}$ and  $\dot S_{{\cal J}}=\lim_{T\rightarrow \infty}\frac{1}{2T} \langle \ln \tilde {\cal J}[{\bf X}]/{\cal J}\rangle_{st}$. Since $\langle R_{cg}[{\bf X}]\rangle_{st}$ is non-negative, Eq. (\ref{Eqasym}) implies that
\begin{align}
\label{Eqbound}
\dot S_m\ge \dot S_{{\cal J}}\ ,
\end{align}
which may be regarded as the generalized second law for the feedback controlled system. This is the central result of this Letter. The contribution  $\dot S_{{\cal J}}$ represents the entropic cost of the   feedback control and can be either negative or positive. It may be interpreted as a phase space `contraction'  or `expansion' induced by the non-standard time-reversal transformation that leads to Eq. (\ref{Eqratio}) (see also the comment below after Eq. (20)). 

In addition to the inequality (\ref{Eqbound}), we conjecture the following asymptotic integral fluctuation relation
\begin{align}
\label{EqIFT2}
\lim_{T\rightarrow \infty}\frac{1}{2T} \ln \langle e^{-\big(\Delta s_{tot}[{\bf X},{\bf X}_-]-\ln \frac{\tilde {\cal J}[{\bf X}]}{{\cal J}}\big)}\rangle_{st}=0 
\end{align}
which is strongly supported by analytical\cite{MRT2014}  and numerical calculations (see Figs. 1 and 2). (Note that Eq. (\ref{EqIFT2}) involves $\Delta s_{tot}$ and not $\Delta s_m$. The latter displays strong fluctuations which are stabilized by the border term.)

{\it Expression of the Jacobian.} The Jacobian  $\tilde {\cal J}[{\bf X}]$ thus plays a central role as the footprint  of non-Markovianity and we devote the rest of this Letter to its calculation.  The starting point is the operator representation of  the conjugate, non-causal Langevin equation. Generalizing the analysis of \cite{OO2007,ABC2010}, one easily finds that $\tilde {\cal J}[{\bf X}]$ can be formally expressed as
\begin{align}
\label{Eqexp}
\tilde {\cal J}[{\bf X}]&={\cal J}\exp \mbox{Tr}\ln[\delta_{t-t'}-\tilde M_{tt'}]\nonumber\\
&={\cal J}\exp-\sum_{n=1}^{\infty}\frac{1}{n}\int_{-T}^{T} dt\: \Big\{\underbrace{\tilde M\circ \tilde M\circ...\tilde M}_{n \: \mbox{times}}\Big\}_{tt}\
\end{align}
where the operator $\tilde M(t,t')$ is defined by
\begin{align}
\label{EqdefM}
\tilde M(t,t')= \{G\circ \tilde F'_{tot}\}_{tt'}\equiv \int_{-T}^{T} dt''\: G(t-t'')\tilde F'_{tot}(t'',t') \ .
\end{align}
$G(t)$ is the Green function for the inertial and dissipative terms in the Langevin equation, and $\tilde F_{tot}^{'}(t,t')\equiv \delta \big[F(x(t))+\tilde F_{fb}[{\bf X},{\bf X}_-]\big]/\delta x(t')$. In the white noise limit, one simply has $G(t)=\gamma^{-1}[1-e^{-\gamma t/m}]\Theta(t)$, where  $\Theta(t)$ is the Heaviside step function\cite{ABC2010}.

{\it Application to linear Langevin processes.--}  To be more specific, let us now consider the case of a harmonic oscillator submitted to a linear feedback control, which is relevant to many practical applications. Since we assume that the noise in Eq. (\ref{EqL1}) is white and Gaussian, all probabilities are  Gaussian in the steady state and thus ${\cal P}[{\bf X}^\dag] ={\cal P}[ {\bf X}]$. As already stressed, this implies that the quantity $\langle \ln {\cal P}[ {\bf X}]/{\cal P}[{\bf X}^\dag]\rangle_{st}$ which  is commonly regarded as a measure of irreversibility (even for non-Markovian processes\cite{GPVdB2008,RP2012,DE2013}) is a misleading indicator, in contrast with the quantity $R_{cg}[{\bf X}]$ introduced above. 

The crucial simplification due to linearity is that  the functional derivative $\tilde F'_{fb}(t,t')$ and thus $\tilde {\cal J}$  become path-independent. In what follows, we  only consider the  behavior for $T\rightarrow \infty$ and defer a more extensive analysis to \cite{MRT2014}. The  operation $\circ$ in Eqs. (\ref{Eqexp})-(\ref{EqdefM}) is then a convolution and $\tilde M(t,t')$ becomes a function of $t-t'$. This implies that  $\ln \tilde {\cal J}/{\cal J}$ is proportional to $2T$, the duration of the trajectory, and the asymptotic  rate $\dot S_{{\cal J}}=\lim_{T\rightarrow \infty}\frac{1}{2T} \ln \tilde {\cal J}/{\cal J}$ is obtained by Laplace transforming Eq. (\ref{Eqexp}),
\begin{align}
\label{Eqratio3}
\dot S_{{\cal J}}&=\frac{1}{2\pi i}\int_{c-i\infty}^{c +i\infty} ds \: \ln [1-\tilde M(s)]\nonumber\\
&=-\frac{1}{2\pi i} \sum_{n=1}^{\infty}\frac{1}{n}\int_{c-i\infty}^{c +i\infty} ds \: [\tilde M(s)]^n
\end{align}
where $\tilde M(s)\equiv \int_{-\infty}^{\infty} dt \: \tilde M(t) e^{-st}$ and  $s=c +i\omega$.
This can be also expressed as
\begin{align}
\label{EqSJ}
\dot S_{{\cal J}}=\frac{1}{2\pi i}\int_{c-i\infty}^{c +i\infty} ds \: \ln \frac{G(s)}{\tilde \chi(s)}
\end{align}
where $\tilde \chi(s)=[G(s)^{-1}-\tilde F_{tot}'(s)]^{-1}=[ms^2+\gamma s-\tilde F_{tot}'(s)]^{-1}$ is the  Laplace transform of the response function $\tilde \chi(t)$ associated with the conjugate Langevin equation. Note that we use here the bilateral Laplace transform because $\tilde \chi(t)$ is non-zero for $t<0$. In general, the integral in Eq. (\ref{EqSJ}) must be computed numerically by properly choosing the value of $c$ (see Supplemental Material).  
\begin{figure}[hbt]
\begin{center}
\includegraphics[width=7cm]{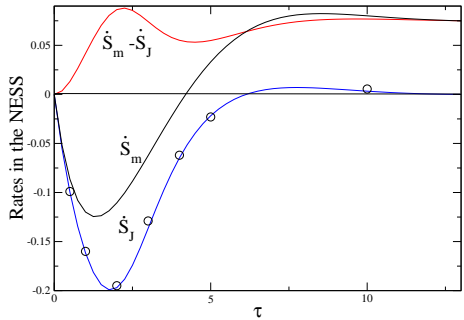}
 \caption{\label{fig1} (Color on line) The rates $\dot S_m$, $\dot S_{{\cal J}}$ and $\dot R_{cg}=\dot S_m-\dot S_{{\cal J}}$ as a function of $\tau$ for the delay Langevin  Eq.  (\ref{EqDL1}) with $m=1,\gamma=1,a=0.5$ and $b=-0.25$. The open circles are obtained from the equation $\dot S_{{\cal J}}\approx -\frac{1}{2T} \ln \langle e^{-\Delta s_{tot}}\rangle_{st}$ using $T=10$ and averaging over $10^6$ independent simulations of Eq. (\ref{EqDL1}) with Heun's method and a time step $\Delta t=10^{-3}$.}
\end{center}
\end{figure}

As a first application, we consider the stochastic delay equation 
\begin{align}
\label{EqDL1}
m\ddot x(t)+\gamma \dot x(t)+ax(t)+bx(t-\tau)=\xi(t) 
\end{align}
which arises in a variety of mechanical or biological systems (e.g. in neural networks involved in the  control of movement, posture, and vision\cite{PFFBT2006}) and has been considered previously in the overdamped limit $m=0$\cite{MIK2009,JXH2011} (see the related discussion in \cite{MRT2014}). When $m\ne 0$, the system settles into a  NESS which is stable in a certain region of the parameter space and is characterized by an effective kinetic temperature $T_k \equiv m\langle\dot x^2\rangle_{st}$\cite{MRT2014}. Then $\dot S_m=\frac{\gamma}{m}(\beta T_k-1)$, which may become negative when the feedback is positive ($b<0$) and cools the system. This indicates that another entropic contribution must be taken into account in order to be consistent with the second law. 

Focusing  on the long-time limit, we first compute  $\dot S_{{\cal J}}$ from expansion (\ref{Eqratio3}) which yields (see Supplemental Material)
\begin{align}
\label{Eqtauexp}
\dot S_{{\cal J}}=\frac{b}{m}\tau-\frac{b\gamma}{2m^2}\tau^2+\frac{b(\gamma^2-am-4bm)}{6m^3}\tau^3+{\cal O}(\tau^4) \ .
\end{align}
Interestingly, if one replaces $b\tau$ by $-\gamma'$, the first-order term identifies with the so-called ``entropy pumping'' rate $\dot S_{pu}=-\gamma'/m$ characteristic of a velocity-dependent feedback control \cite{KQ2004,MR2012}. One indeed recovers a force proportional to the velocity by expanding $x(t-\tau)$ at first order in $\tau$. In this sense, $\dot S_{{\cal J}}$ may  be viewed as a generalization of $\dot S_{pu}$. To go beyond the small-$\tau$ expansion,  Eq. (\ref{EqSJ}) must be integrated numerically, using $\tilde \chi(s)=[ms^2+\gamma s+a+be^{s\tau}]^{-1}$.

As an illustration, we plot in Fig. 1 the rates $\dot S_m$, $\dot S_{{\cal J}}$, and $\dot R_{cg}=\dot S_m-\dot S_{{\cal J}}$  as a function of $\tau$ in the case of a positive feedback. One can see that $\dot R_{cg}$ is always positive, in agreement with the generalized second law, Eq. (\ref{Eqbound}).
The non-monotonous behavior of  $\dot S_m$ is directly dictated by the behavior of $T_k$, which is not the case for $\dot S_{{\cal J}}$. Note also that  $\dot S_m$ goes to a finite value for $\tau \rightarrow \infty$ whereas  $\dot S_{{\cal J}} \rightarrow 0$.
 We also indicate in the figure some values of $\dot S_{{\cal J}}$ obtained by simulating the Langevin equation (\ref{EqDL1}) and using Eq. (\ref{EqIFT2}) which takes the simple form $\lim_{T\rightarrow \infty}\frac{1}{2T} \ln \langle e^{-\Delta s_{tot}[{\bf X},{\bf X}_-]}\rangle_{st}=-\dot S_{{\cal J}}$ for a linear system. As can be seen, the agreement with the theoretical value is already very good with $T=10$. 

As second application, we consider the equation
\begin{align}
\label{Eqvelo}
m\ddot x+\gamma \dot x+ax+\frac{\gamma'}{\tau}\int_{-\infty}^{t} dt' \: e^{-\frac{t-t'}{\tau}}\dot x(t')= \xi(t)
\end{align}
which may describe a feedback-cooled electromechanical oscillator\cite{PZ2012,GRBC2009}. The molecular refrigerator model of \cite{KQ2004,MR2012} is recovered  in the Markovian limit  $\tau \rightarrow 0$.  
Since the system is linear, this  also amounts to studying the coupled Markovian equations\cite{MR2013}
\begin{align}
m\ddot x+\gamma \dot x+ax+\gamma' y&= \xi(t)\nonumber\\
\dot y+\frac{1}{\tau}(y-\dot x)&=\eta (t)
\end{align}
in the limit where the  noise $\eta$  becomes negligible. More generally, such coupled equations are  useful to investigate the role of coarse-graining and hidden degrees of freedom on fluctuation theorems\cite{CPV2012,KN2013,IS2013}. 

\begin{figure}[hbt]
\begin{center}
\includegraphics[width=7cm]{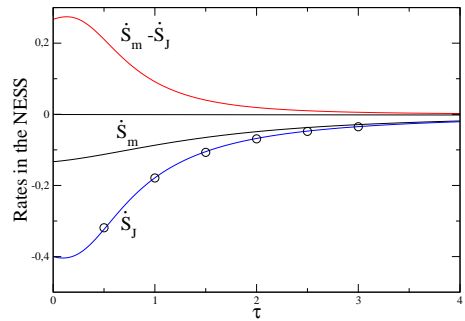}
 \caption{\label{fig2} (Color on line) Same as Fig. 1 for the velocity-dependent feedback described by Eq. (\ref{Eqvelo}). The model parameters are $m=1, a=1,\gamma=0.2,\gamma'=0.4$. Note that $\dot S_{{\cal J}}\rightarrow -\frac{\gamma'}{m}$ for $\tau \rightarrow 0$.}
\end{center}
\end{figure}
For $\gamma'>0$, heat  permanently flows from the bath to the system in the steady state, with a rate given by Eq. (77) in \cite{MR2013} with $T'=0$. This yields
$\dot S_m =-(\gamma \gamma')/(m\gamma_{eff})$ where $\gamma_{eff}=(\gamma +\gamma')(1+\gamma \tau/m)+a\gamma \tau^2/m$.  The conjugate dynamics is now  defined by the changes $\tau \rightarrow -\tau$ and $\gamma' \rightarrow -\gamma'$, and the expansion (\ref{Eqratio3})  then yields 
\begin{align}
\dot S_{{\cal J}}=-\frac{\gamma'}{m}+\frac{\gamma'(\gamma-\gamma')}{m^2}\tau+{\cal O}(\tau^2)\ .
\end{align}
As it must be, the first term is just the entropy pumping contribution obtained in \cite{KQ2004} in the Markovian limit. This demonstrates that the present formalism  is valid for both position- and  velocity-dependent feedback control.

 Some typical results for the rates as a function of $\tau$ are shown in Fig. 2. One again observes that the generalized second law (\ref{Eqbound}) is obeyed and that Eq. (\ref{EqIFT2}) is in good agreement with the numerical simulations of the Langevin equation. In this model, both $\dot S_m$ and $\dot S_{{\cal J}}$ go to zero as $\tau \rightarrow \infty$.

{\it Summary} -- By studying the nature of time-reversal breaking in the action functional of the path space measure,  we have identified the unusual mathematical mechanism that contributes to the positivity of the entropy production in Langevin systems submitted to a continuous (position- or velocity-dependent) non-Markovian feedback control.  In particular, the present formalism extends the framework of stochastic thermodynamics to the vast class of time-delayed diffusion processes. 
An important step further will be to include measurement noise. This will also clarify the relationship with previous approaches, in particular the abstract theoretical setup presented in \cite{SU2012}, which still remains elusive.

\vspace*{7mm}

We are grateful to G. Tarjus for his help in the interpretation of the IFT.  M.L.R.  also acknowledges useful exchanges with S. Ito and T. Sagawa. 

\vspace*{7mm}

\textbf{Supplemental Material: Computation of $\dot S_{\cal J}$ for linear systems}

As indicated  in the main text, the asymptotic rate $\dot S_{\cal J}$ is conveniently computed in linear systems  by taking the bilateral Laplace transform of Eq. (15). We give here some more details about the calculation. 

The starting point is Eq. (16) which defines $\tilde M(t)$ for $T\rightarrow \infty$ as the convolution
\begin{align}
\tilde M(t)\equiv \int_{-\infty}^{\infty} dt' \: G(t-t') \tilde F_{tot}'(t'-t)\ .
\end{align}
This yields  $\tilde M(s)\equiv \int_{-\infty}^{\infty} dt \: \tilde M(t) e^{-st} =G(s)\tilde F_{tot}'(s)$ (with $s=c+i\omega$), and thus $\tilde \chi(s)\equiv G(s)[1-\tilde M(s)]^{-1}=[G(s)^{-1}-\tilde F_{tot}'(s)]^{-1}$, where 
\begin{align}
G(s)=\frac{1}{\gamma}\int_{0}^{\infty} dt (1-e^{-\gamma t/m})e^{-st}=\frac{1}{ms(s+\gamma /m)} \ .
\end{align}

We first consider the model governed by the time-delayed Langevin equation (19). The conjugate dynamics is  defined by the change $\tau \rightarrow -\tau$,  so that $\tilde F'_{tot}(t)=-a\delta(t)-b\delta (t+ \tau)$. This yields
\begin{align}
\tilde M(t)&=-\frac{a}{\gamma} (1-e^{-\gamma t/m})\Theta(t)\nonumber\\
&-\frac{b}{\gamma} (1-e^{-\gamma (t+\tau)/m})\Theta(t+\tau)
\end{align}
and thus $\tilde M(s)=-[a+be^{s\tau}][ms(s+\gamma/m)]^{-1}$ with the region of convergence (ROC) defined by $c>0$. In order to compute the series expansion defined in the second line of Eq. (17), the integration contour  is then closed to the left-hand side of the complex $s$-plane and the successive terms in the series are  obtained  by adding the residues at the two poles $s=0$ and $s=-\gamma/m$. After reordering\cite{MRT2014}, this yields the series expansion in $\tau$ given by Eq. (20). On the other hand, choosing the value of $c$ for the contour integral in Eq. (18) is a more delicate issue which requires to determine the position of the poles of $\tilde \chi(s)=[ms^2+\gamma s+a+be^{s\tau}]^{-1}$. A careful analysis shows that $\tilde \chi(s)$ has two poles on the l.h.s. of the  complex $s$-plane and an infinity of poles on the r.h.s., which is the signature of non-causality. One can show\cite{MRT2014} that the correct choice for the integration contour is $0<c<c_1^+$, where $c_1^+$ denotes the pole closest to the imaginary axis. The numerical integration of Eq. (18) is then in agreement with the series expansion (20) when the latter converges.

We next consider the  model described by Eq. (21) in the main text. The conjugate dynamics  is  defined by the changes $\tau \rightarrow -\tau$ and $\gamma' \rightarrow -\gamma'$ so that  
$\tilde F'_{tot}(t)=-[a+(\gamma'/\tau)]\delta (t)+(\gamma'/\tau^2) e^{t/\tau}\Theta (-t)$ by partial integration of $\tilde F_{fb}(t)$.  Hence 
\begin{align}
\tilde M(t)&=\left[-\frac{a}{\gamma} (1-e^{-\gamma t/m})+\frac{\gamma'/m}{1+\gamma\tau/m}e^{-\gamma t/m}\right]\Theta(t)\nonumber\\
&+\frac{\gamma'/m}{1+\gamma\tau/m}e^{t/\tau} \Theta (-t),
\end{align}
and $\tilde M(s)=[a-(a\tau+\gamma')s]/[ms(s+\gamma/m)(\tau s-1)]$ with the ROC defined by $0<c<1/\tau$. The expansion  (23) is then obtained from Eq. (17) by closing the contour either to the l.h.s and adding the residues at $s=0$ and $s=-\gamma/m$   or  to the r.h.s. and taking the residue at $s=1/\tau$. To perform the numerical  integration in Eq. (18), one must  consider the position of the poles of
$\tilde \chi(s)=[\tau s-1][m\tau s^3 +(\gamma \tau -m)s^2+(\gamma' -\gamma +a\tau)s-a]^{-1}$. In particular, the situation depends on the sign of $\gamma'-\gamma$, which is  reminiscent of the behavior of the large deviation function for the entropy production studied in \cite{MR2012}.

\end{document}